\documentclass[conference]{IEEEtran}
\usepackage{blindtext}
\usepackage{graphicx}
\usepackage{color}
\usepackage{gensymb}

\ifCLASSINFOpdf
\else
\fi
\begin{document}
%
\title{Real-Time Index Authentication for Event-Oriented Surveillance Video Query using Blockchain}


%
%
%

\author{
\IEEEauthorblockN{Seyed Yahya Nikouei, Ronghua Xu, Deeraj Nagothu, Yu Chen}
\IEEEauthorblockA{
\textit{Binghamton University, SUNY} \\ Binghamton, NY 13902, USA \\
\{snikoue1, rxu22, dnagoth1, ychen\}@binghamton.edu}
\and
\IEEEauthorblockN{Alexander Aved, Erik Blasch}
\IEEEauthorblockA{\textit{The U.S. Air Force Research Laboratory} \\
Rome, NY 13441, USA \\
\{alexander.aved, erik.blasch\}@us.af.mil}
}

\maketitle

\begin{abstract}

Information from surveillance video is essential for situational awareness (SAW). Nowadays, a prohibitively large amount of surveillance data is being generated continuously by ubiquitously distributed video sensors. It is very challenging to immediately identify the objects of interest or zoom in suspicious actions from thousands of video frames. Making the big data indexable is critical to tackle this problem. It is ideal to generate pattern indexes in a real-time, on-site manner on the video streaming instead of depending on the batch processing at the cloud centers. The modern edge-fog-cloud computing paradigm allows implementation of time sensitive tasks at the edge of the network. The on-site edge devices collect the information sensed in format of frames and extracts useful features. The near-site fog nodes conduct the contextualization and classification of the features. The remote cloud center is in charge of more data intensive and computing intensive tasks. However, exchanging the index information among devices in different layers raises security concerns where an adversary can capture or tamper with features to mislead the surveillance system. In this paper, a blockchain enabled scheme is proposed to protect the index data through an encrypted secure channel between the edge and fog nodes. It reduces the chance of attacks on the small edge and fog devices. The feasibility of the proposal is validated through intensive experimental analysis. 
\end{abstract}


%
\IEEEpeerreviewmaketitle

\section{Introduction}
\label{sec:intro}
Thanks to the proliferation of the Internet of Things (IoT) technology that links cyber-physical systems and social objects, the concept of Smart Cities becomes feasible and provides high-value services that improve the life quality of its residents. As one of the most actively researched smart city topics, smart surveillance enables a broad spectrum of promising applications, including access control in areas of interest, human identity or behavior recognition, crowd flux statistics and congestion analysis, detection of anomalous behaviors, and interactive surveillance using multiple cameras \cite{henriques2015high}. Information from surveillance video is essential to achieve situational awareness (SAW). A prohibitively large amount of surveillance data is being generated continuously every second by the ubiquitously distributed video sensors. Owning to the onerous computation requirement of big data contextual tasks, many of smart surveillance applications rely on a centralized cloud computing framework that possesses abundant computation power, excellent flexibility, and scalability. The massive amount of raw frame data have to be transferred to cloud centers, however, it inevitably incurs uncertain latency and poses extra workload to the communication networks. Furthermore, it is very challenging to immediately identify the objects of interest or zoom in suspicious actions from thousands of video frames. Generating pattern indexes in a real-time, on-site, and over big data manner on the video streaming instead of depending on the batch processing at the cloud centers is a strategical advantage for system deployment \cite{aved2015multi}.


To address the challenges in smart surveillance system running on cloud-based architecture, fog/edge computing has been recognized as a promising approach that migrates computation tasks to the edge of a network. Merging more intelligence to the ubiquitously deployed networked cameras and smart mobile devices allows more jobs conducted by the decentralized nodes at the edge of networks. It enables the smart surveillance system to meet the delay-sensitive, mission-critical requirements \cite{chen2017enabling}, \cite{nikouei2018lcnn}, \cite{xu2018real}. The distributed edge/fog devices locally process raw video streams and makes the video indexable by extracting, recognizing. and labeling useful features. The feature description and index data are transferred to the nodes in higher layer to help advanced analytic tasks. However, the remote data transmission also incurs concerns in data security and privacy because it expose vulnerabilities to potential attackers to perform malicious operations, such as Denial of Service (DoS) attacks, false video injection attacks, modifying tracking trajectory, and eavesdropping private video streams.  

To deter the attackers from hijacking the data from the channel between the edge devices and fog nodes, it is necessary to build a secure communication channel, which consists of four steps: \textit{encrypting, sending, receiving and decrypting}. To break this secure channel an attacker would have to intercept the channel and decrypt the data. Due to the limited computational power available at the edge devices, however, the encryption is relatively weak. Once the encryption scheme is known to the attacker, it is not difficult to gain access to the data \cite{daemen2013design}. An ideal solution is a hybrid encryption mechanism leveraging both the symmetric and asymmetric key encryptions.

Since the smart surveillance system is deployed in a distributed network environment including an extraordinary large number of IoT devices with high heterogeneity and dynamics, it necessitates a more scalable, flexible and lightweight security mechanism for the distributed IoT network. Furthermore, those smart devices are geographical scattered across near-site edge network in an untrusted network environment. It is not suitable to enforce security on a centralized authority, which suffers from the performance bottleneck or the single point of failure. Thus, the smart surveillance system needs a new decentralized framework that provides security schemes in the trust-less application network environments.

As the fundamental protocol of Bitcoin \cite{nakamoto2008bitcoin}, which was the first digital currency, the blockchain protocol has been recognized as the potential to revolutionize the fundamentals of IT technology owing to its many attractive features and characteristics such as supporting decentralization and anonymity maintenance \cite{ouaddah2016fairaccess}. In this paper, a blockchain enabled index authentication scheme for real-time event-oriented surveillance video query is proposed to enhance the security of smart surveillance system. Through executing detection and tracking tasks on the embedded edge devices, event-oriented surveillance service extracts featured information by processing input frames. Then, a real-time indexing service generates unique index for each frame to prevent malicious modification on image. Finally, the frame indexes are imprinted to a blockchain network and verified by a decentralized smart contract based authentication mechanism. Experimental results demonstrate the feasibility and effectiveness of the proposed scheme.

The major contributions of this work are:

\begin{itemize}
\item[1)] A complete architecture of real-time index authentication scheme for smart surveillance system is proposed, which includes event-oriented surveillance video query, real-time indexing, and blockchain-enabled authentication;

\item[2)] A proof of concept prototype based on smart contracts is implemented and deployed on a local private blockchain network; and

\item[3)] A comprehensive experimental study has been conducted, and the experimental results validate the feasibility of the proposed scheme in IoT environments without introducing significant overhead.
\end{itemize}

The remainder of this paper is organized as follows: Section \ref{sec:back} analyzes and reviews the state of the art research in smart surveillance and blockchain on IoT system. Section \ref{sec:RTindex} illustrates the details of the proposed real-time index authentication strategy. Beside the implementation of the proof-of-concept prototype, Section \ref{sec:experiment} reports an extensive experimental study using test scenarios that are executed on both resource-constrained and non-resource-constrained devices. Finally, a summary is presented in Section \ref{sec:conclusion}.


\section{Background and Related Work}
\label{sec:back}

\subsection{Smart Surveillance at the Edge}
\label{subsec:background surveillance}
Most surveillance systems that are available today for purchase will function as an archive of footages and being used for off-line forensics analysis and depends on human operators in process loop \cite{chamasemani2013systematic}. Implementation of the on-line surveillance tasks are limited because of the time delay and uncertainties associated with the approach of sending footage to distance servers. On the other hand, recent Machine Learning (ML) algorithms require less powerful processors for operation and promising results are achieved using them at the edge of the network \cite{nikouei2018intelligent}. Such that human detection and tracking along with abnormal behavior detection analysis are feasible at edge or fog computing layers using various smart deep learning or other ML algorithms. In most cases the video streams are transferred to a cloud node for further processing \cite{vishwakarma2013survey} that creates a burden on the communication network. 

The smart surveillance community has recognized that heavy communication overhead is not tolerable in many mission-critical, delay sensitive tasks \cite{chen2018smart}, \cite{shi2016edge}. Recent developments of the edge devices and their appropriate hierarchy models enable real-time surveillance leveraging the fog computing paradigm \cite{mahmud2018fog}. Based on the fog computing paradigm, on-line and uninterrupted target tracking systems are proposed to meet the requirements of real-time video processing and instant decision making \cite{mukherjee2018survey}. For example, researchers have tried to merge raw video stream generated by drones on near-site fog computing devices, such as tablet or laptop \cite{chen2016dynamic}.

\begin{figure}[t]
    \centering
        \includegraphics[width=0.425\textwidth]{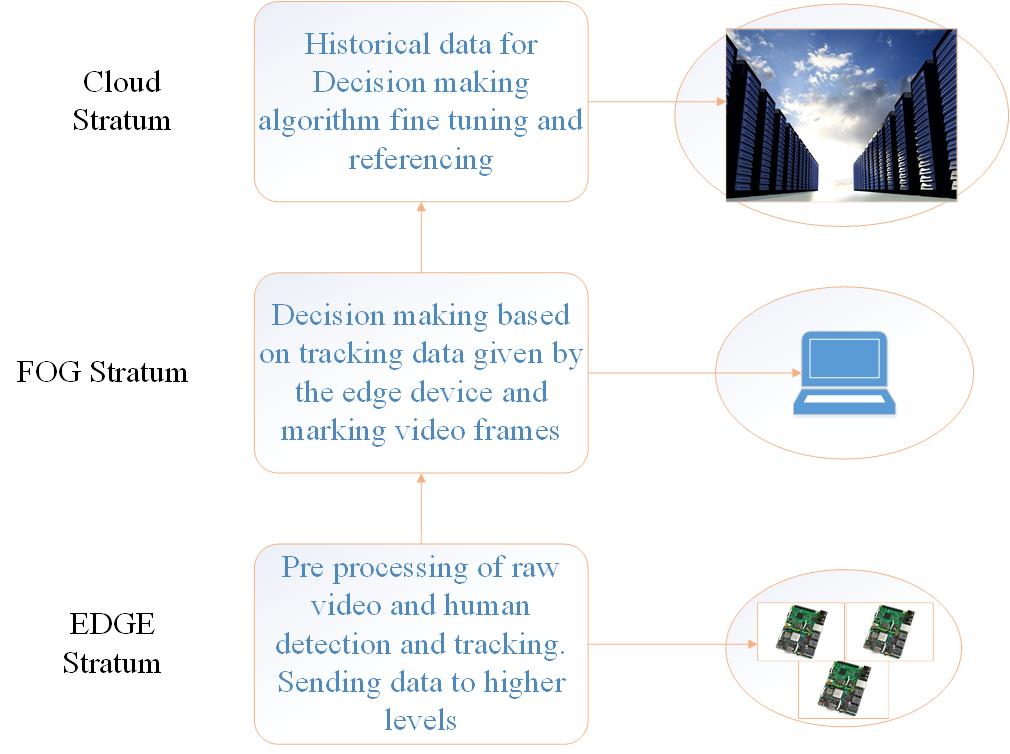}
    \caption{Layered smart surveillance system hierarchy in the edge-fog-cloud computing paradigm.}
    \label{fig:implementation}
    \vspace{-10pt}
\end{figure}

The surveillance system that is focused on human objects conducted at the edge can be constructed following the edge hierarchy architecture shown in Fig. \ref{fig:implementation}, which includes edge, fog, and cloud stratum \cite{mouradian2017comprehensive}. Analytically, a smart surveillance task can be considered as a three-layer framework. In the first step the input frame of the surveillance camera is given to an edge device and the low-level information is extracted, which is not too heavy for edge device implementation, such as feature detection and object tracking  \cite{howard2017mobilenets}, \cite{xu2018real}. The intermediate-level is in charge of mode recognition like action recognition and behavior understanding, and decision making like abnormal event detection, which is implemented at the fog stratum. Finally, the high-level is focused on fine tunning of the algorithms, historical profile building, and global statistical analysis.

Connections among the edge nodes, fog nodes and cloud servers present challenges in terms of quality of service (QoS) and security. The video frame processing tasks are computing intensive and outsourcing of some functions is inevitable. While sending the frames to a distant cloud takes time, the near-site fog nodes are powerful enough for abnormal behavior detection. In a smart transportation application, such a hierarchical system allows read data from the sensors implemented on buses and then transfers them to a fog node where contextualization takes place and the data is classified \cite{cao2017edge}. 

\subsection{Secure Communication Channel}
\label{subsec:encrypt-decrypt}

Symmetric encryption includes a secret key which can be a number, word or just a string based on user preference. The key is shared only between the trusted sender and receiver that makes it possible for only these two nodes to encrypt and decrypt the text. The problem with symmetric encryption is that the key could fall into wrong hands while sharing via the Internet, and anyone with the key can decrypt the text.

Asymmetric encryption includes a key pair known as the public and private key. 
The text encrypted with the public key can only be decrypted with user’s private key which is only known by the user. The private key provides no other key sharing through the Internet. The trade-off with asymmetric encryption is that it requires far more processing power and it is slower than symmetric encryption. Therefore, in order to leverage the advantages from both of the encryption techniques, a hybrid solution is promising. In this work, we choose the AES (Advanced Encryption standard) and RSA (Rivest–Shamir–Adleman)
 encryption algorithm as the symmetric and asymmetric key encryption algorithms \cite{yu2018survey}.

Previously, the encryption techniques like AES and RSA have been used in cloud computing security and image steganography to enhance the security performance. AES has been adopted to generate random pixel position as well as to decide the size of least significant bits for embedding information dynamically \cite{lalengmawia2016image}. Cryptography and steganography are used simultaneously for data encryption \cite{ebrahim2017hybrid}, in which the hash of the data and AES encrypted key is encrypted with public key encryption. This encrypted data is embedded in images using the least significant bit technique in steganography. A similar technique was proposed that uses a combination of AES and RSA in secure cloud systems \cite{khanezaei2014framework}, which improves the information transmission performance between the user and the cloud data storage.

\subsection{Blockchain and Smart Contract in IoT}
Blockchain is a fundamental technology of Bitcoin, which was introduced by Nakamoto in 2008 \cite{nakamoto2008bitcoin}. The blockchain is a public ledger that provides a verifiable, append-only chained data structure of transactions. Through allowing the data be stored and updated distributively, the blockchain is essentially a decentralized architecture that does not rely on a centralized authority anymore.  The transactions are approved by miners and recorded in timestamped blocks, where each block is identified by a cryptographic hash and chained to preceding blocks in a chronological order. Blockchain uses \textit{consensus mechanism}, which is enforced on a large amount of distributed nodes called miners, to maintain the sanctity of the data recorded on the blocks. Thanks to the “trustless” proof mechanism running on miners across networks, users can trust the system of the public ledger stored worldwide on many different decentralized nodes maintained by ''miner-accountants'', as opposed to having to establish and maintain trust with a transaction counter-party or a third-party intermediary \cite{swan2015blockchain}. Thus, Blockchain is an ideal decentralized architecture to ensure distributed  transactions between all participants in a trustless environment, like IoT networks.

Because of so many attractive characteristics, researchers have verified the feasibility of applying blockchain technique to address security issues in the IoT networks, such as access control \cite{maesa2017blockchain}, \cite{ouaddah2016fairaccess}. Blockchain has shown its success in decentralization of currency and payments, like Bitcoin. Designing programmable contracts that support variety of flexible transaction types becomes a trend to extend blockchain's applications beyond cryptocurrency. Emerging from the smart property, smart contract allows users to achieve agreements among parties through blockchain network instead of relying on third parties to maintain a trust relationship. By using cryptographic and security mechanisms, smart contract combines protocols with user interfaces to formalize and secure relationships over computer networks~\cite{szabo1997formalizing}. \textit{Smart contract} includes a collection of pre-defined instructions and data that have been saved at a specific address of blockchain as a Merkle hash tree, which is a constructed bottom-to-up binary tree data structure. Through exposing public functions or application binary interfaces (ABIs), smart contract interacts with users to offer predefined business logic or contract agreement. Smart contract enabled security mechanism for IoT systems has been a hot topic and some efforts have been reported recently, for example, data protection \cite{lee2018implementation} and access control \cite{xu2018blendcac,xu2018smartcac}, \cite{zhang2018smart}. Blockchain and smart contract together are promising to provide a solution to enable index authentication in distributed smart surveillance systems.

\begin{figure*}[t]
    \centering
        \includegraphics[width=0.67\textwidth]{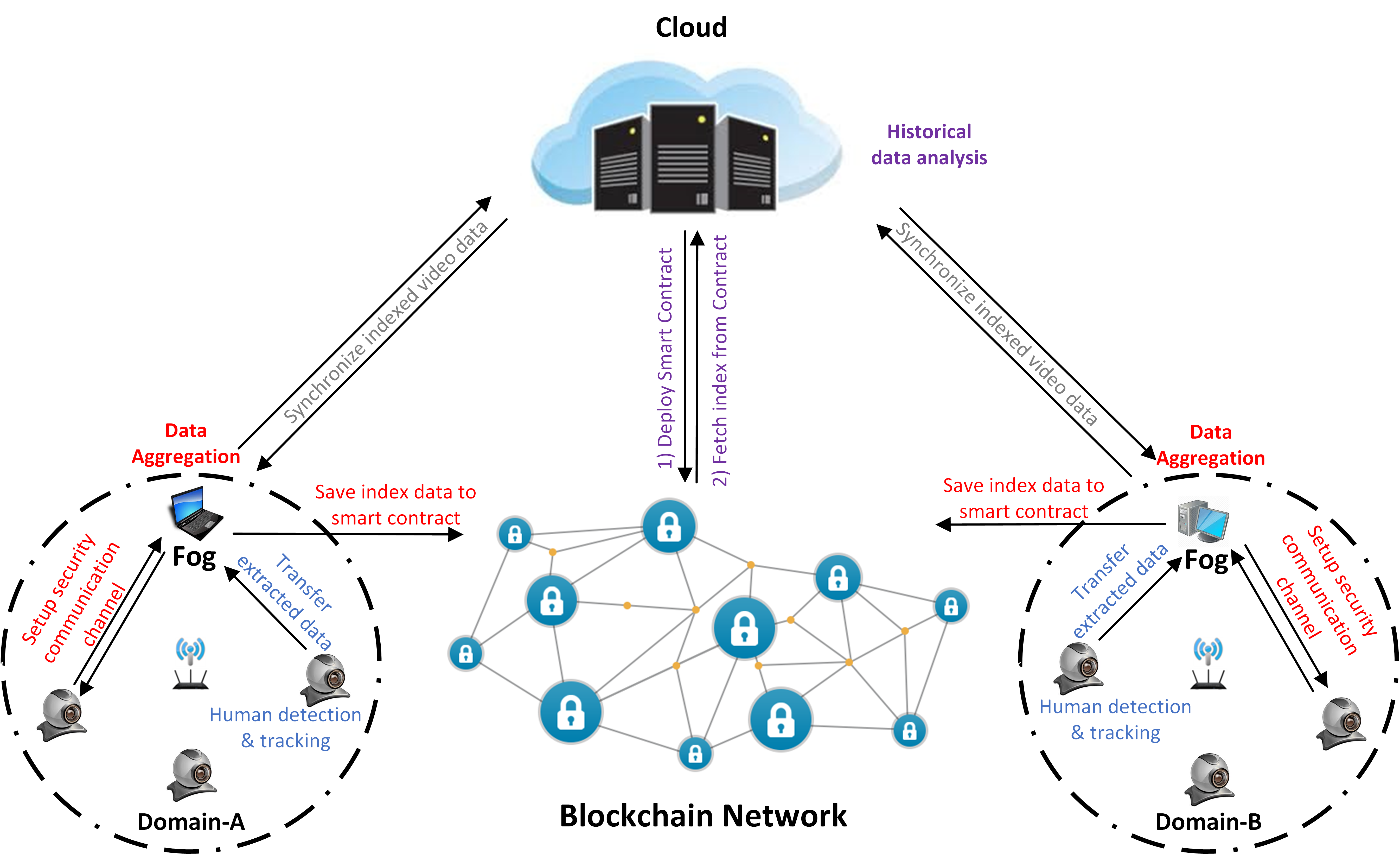}
    \caption{Illustration of the System Architecture.}
    \label{fig:framework}
    \vspace{-10pt}
\end{figure*}

\section{Real-Time Index Authentication}
\label{sec:RTindex}

Inspired by the smart contract and blockchain technology, a real-time index authentication for event-oriented surveillance video query system is proposed to provide a decentralized video streams security mechanism in the untrusted edge network environment. Figure \ref{fig:framework} illustrates the proposed system framework, which demonstrates a scenario including two isolated IoT-based video surveillance domains without a pre-established trust relationship. With object detection and tracking tasks conducted by the smart cameras, low-level feature information is extracted on-site by processing surveillance video streaming at the network edge, then transferred to fog devices for data aggregation and further analysis. In each domain, the fog device not only enforces predefined security policies to manage domain related devices and services, but also acts as an intermediate to interact with public blockchain and cloud to enable the index authentication for event-oriented surveillance video query. The main components of the framework include the event-oriented surveillance video query, real-time indexing and secure data transferring, and blochchain-enabled authentication.

\subsection{Smart Surveillance System and Secure Data Transfer}
\label{sec:smart system}

Processing the video instantly gives better understanding of the event taking place in real time. The surveillance camera captures the video and transfers it to the edge/fog devices of choice in real-time. The edge device is connected through Local Area Network (LAN) to the camera and is located on site. It takes each frame as the first point of the automated abnormal detection mechanism. 

After the reception of the frame, the edge device is in charge of extracting low-level features for abnormal behavior detection. In order to have a functional system for anomalous behavior detection or prediction, the surveillance system needs to accurately identify objects. Otherwise the system may miss anomalies or incur a high false alarm rate. Based on oblique human movement and change of appearance, pedestrian detection using less recourses still is considered a challenging question in computer vision \cite{howard2017mobilenets}. Also, in the application of the smart surveillance the edge device that is responsible for human object detection has constraints on the computing power and storage resources available for this job. State of the art architectures of Convolutional Neural Networks (CNN) are widely used in recent years for object detection or classification, we have witnessed the efforts in designing lightweight classifiers for edge and mobile device implementations \cite{nikouei2018lcnn}. 

Running a person, object, vehicle (POV) detection algorithm proves to be heavy for the edge devices in practice and so a more lightweight tracker may follow the object after initial detection, while the detection is executed with a extended period. Such that the tracker should be trained on-line using the objects of interest, because of the divergence of the appearance of human objects and the ambient lighting conditions of the scene. Trackers that are introduced in recent years working with pre-trained CNN models such as GoTURN \cite{held2016learning} and SFC \cite{bertinetto2016fully} require a lot of resources that makes them unfit for the applications at the edge. Using a fast and reliable tracker such as Kernelized Correlation Filter (KCF) \cite {henriques2015high} with on-line tracking, grantees real-time tracking. Many surveillance cameras in use today are at 10 frames per second (FPS). In practice the normal velocity of pedestrians is slow, thus even a speed of 5 FPS can be considered acceptable.

The second step in the process is to extract features of the detected objects. The features might include speed, direction and some other descriptive metrics such as certain specific gestures an object may have. Gestures are defined with detection of the head, shoulder, upper arm and lower arm of the object and the angles they may produce. A CNN can classify parts of the body. Current location of people who are detected are kept along their other features as separate objects. 

The edge node will outsource the remaining steps to the fog nodes or cloud center, such as contextualization of the features, classification of the features into normal or abnormal instances, and saving information for future referencing. Once the features are extracted, transferring information between nodes requires security measures to ensure the confidentiality and integrity.

\begin{figure}[t]
    \centering
        \includegraphics[width=0.385\textwidth]{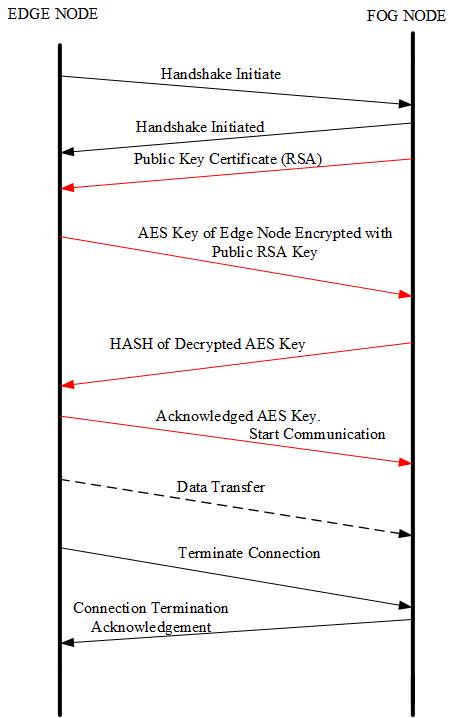}
    \caption{Establishment of double encrypted communication between nodes.}
    \label{fig:encryption_flow}
    \vspace{-10pt}
\end{figure}

The data transfer from the edge node to the fog node is carried out in a secure communication channel encrypted with AES and RSA algorithms. The advantage of using both encryption algorithm is that it provides a short key establishment time and is more robust to the network sniffing attacks. The shared key encryption can be established without the attacker being able to intercept the key. The fog node's public key is used to encrypt the shared key and this encrypted data is used to send the shared key to the fog node to establish the secure shared key channel. The hashes of the shared key is exchanged to verify that the key has been established. The impact on the efficiency of the channel is very low and we can establish double layer encrypted channel. The establishment procedure of the secure channel is demonstrated in Fig. \ref{fig:encryption_flow} and described in the following steps,

\begin{itemize}
\item[1:] The edge node initiates the handshake with the fog Node.

\item[2:] The fog node responds to the handshake and replies with the public key certificate.

\item[3:] The edge node uses the public key of the fog node and sends the encrypted shared key.

\item[4:] The fog node uses its private key and decrypts the shared key received. Then it sends back the computed hash of the shared key.

\item[5:] The edge node verifies the hash of the shared key and sends an acknowledgment. With this, a secure connection is established and the data transfer begins.

\item[6:] Once the Data transfer is finished, the connection is terminated and the shared key is discarded for no future use. A new connection will be established with a new shared key if necessary.
\end{itemize}

\subsection{Real-Time Indexing and Event-Oriented Video Query}
\label{sec:index}

The edge device may not be used for further data processes purposes as the object detection and tracking have consumed most of the resources, as to be illustrated in Section \ref{sec:experiment}. The features extracted from each frame are encrypted and sent to the fog node, where it will be decrypted and used for contextualization, which refers to the act of placing the features in a spatio-temporal context. For example, a person walks in the hall ways of the university office area during business hours is normal, but the same activity in the late night might be suspicious. Other than its use in classification, contextualization of the data before storing is a key for search of activity or event of interest in video stream, such as video clips with a person captured during a time of concern.

Therefore, the geo-location of the camera, time or sequence of the frame, total number of the objects in the frame, and their gestures are recorded as matrices in a Key-Value manner. In each given frame, each object has Keys and each callable key has a value assigned to it. These information will be stored at the fog node where storage capacity is available and can be used for future search based on the keys. In practice, these keys are called indexing data used for faster search for information of interest. Similarly, big data storage management systems use the same approach in parallel to search for a keyword value which makes the search faster \cite{shafer2010hadoop}.

Querying the video can be done using the index table. This feature is extremely useful when searching for a specific incident or activity of interest in the video stream. The common practice is to look at the footage slowly and find the moment of interest that in most cases will take considerable amount of time. Using the index table of features will efficiently reduce the search time of query video by querying for a string variable instead of looking into old video files. Once the contextualized data has been saved in the fog node, the target video clips will be indexed using the keywords based on the time, location or other attributes of interest. This approach can provide more powerful searching functions. For instance, once the time and the camera ID is known, the speed of the objects at that scenario may be the query. 

As discussed in Section \ref{sec:smart system}, the data transfer between the edge device and the fog node is through a secured communication channel. When the features for each frame are decrypted to plain text and contextualized, they are stored and their hash value is calculated. In Section \ref{sec:blockchain}, the hash value will be used to construct a smart contract such that the data in the table is protected from being tempered by attackers.

\subsection{Blockchain-enabled Authentication}
\label{sec:blockchain}

The feature information extracted by edge devices and contextual data are merged at the fog layer, which shares information with cloud for high-level tasks. A blockchain-enabled indexing authentication strategy is proposed to enable a decentralized, scalable and secured data sharing service. The key components and  operation are listed as follow:

\begin{enumerate}
\item \emph{Registration}: In a blockchain system, every entity must create at least one main account defined by a pair of keys to join the network. Each account`s address is derived from his/her own public key. Given the scenario shown in Fig. \ref{fig:framework}, the identity authentication and management functions are implemented on the cloud server using the unique account address as the Virtual Identity (VID), which is saved in a global profile database maintained by the cloud. Each fog node could send registration requests to the cloud. Once the identity information related to fog node is verified, the profile of each registered entity is created using its address for authentication process when transferring hashed index table data to a smart contract.

\item \emph{Smart Contract Deployment}: A smart contract, which manages the hashed index table data, must be developed and deployed on the blockchain network by the index authentication policy owner. In our framework, the cloud acts as the data owner and the policy maker who is able to deploy smart contract encapsulating index authentication policy. After the smart contract has been successfully deployed on the blockchain network, it becomes transparent to the whole network. Transparency implies that all nodes in the blockchain can access the transactions and smart contracts recorded in the chain data. Cryptographic and security mechanism provided by the blockchain network protect all the algorithmically specifiable protocols and relationships in the smart contract from malicious interference by third parties in a trustless network environment. Each node readily access all the transactions and the recent state of the smart contract by referring the local synchronized chain data, and interacts with the smart contract through the address and public Remote Procedure Calls (RPC) interfaces.

\item \emph{Hashed Index Record Generation}: To successfully save the hashed index record to the blockchain, a fog node initially sends an access request to the cloud to get a permission for executing the hashed index record generation ABI functions of a smart contract. Given the registered entity information established in the profile database, a policy decision making module evaluates the request by enforcing the predefined authorization policies. If the access request is granted, the cloud launches a transaction to update the authorized entity list in the smart contract. After the transaction has been approved and recorded in a new block, the cloud notifies the fog node with a smart contract address and the ABI function for recording hashed index data. Each time when those hashed index records are available on authorized fog devices, they just simply interact with the authorized ABI function to update the hashed index data on the blockchain.

\item \emph{Index Authentication}: The index authentication process is performed by the authorized entities who are the users of video query services, for example, the cloud nodes in our scenario. When the cloud operator wants to authenticate a video querying data saved on the fog nodes, he/she just simply checks the current state of the contract in the regularly synchronized local chain data to get a hashed key-value index record for proof. Through a comparison between calculated hash value of the record index table and the hashed index record in the blockchain, the cloud operator can verify whether or not the received video query data is authenticate.
\end{enumerate}

\section{Experimental Study}
\label{sec:experiment}

A proof of concept prototype system has been implemented in a real physical network environment to validate the feasibility of the proposed system. Using Solidity \cite{solidity}, which is a contract-oriented, high-level language for smart contracts development, the blockchain-enabled index authentication mechanism has been transcoded to a smart contract and deployed on a real private Ethereum \cite{ehtereum} blockchain network. The hashed index verification function is implemented as a web service application based on the Flask framework \cite{flask} using Python.

\subsection{Testbed Setup}
The edge devices are two Asus Tinker Boards with the configuration as follows: 1.8 GHz 32-bit quad-core ARM Cortex-A17 CPU, the memory is 2GB of LPDDR3 dual-channel memory and the operating system is the TinkerOS based on the Linux kernel. The fog layer functions are implemented on a laptop, in which the configuration is as follows: the processor is 2.3 GHz Intel Core i7 (8 cores), the RAM memory is 16 GB and the operating system is the Ubuntu 16.04. The private Ethereum network includes four miners, which are distributed to four desktops that are empowered with the Ubuntu 16.04 OS, 3 GHz Intel Core TM (2 cores) processor and 4 GB memory. Each miner uses two CPU cores for mining task to maintain the private blockchain network. The data transfer between the fog node and the miner is carried through through an Encryption channel. We used python based socket programming language for both ends of the channel.

\subsection{Performance Evaluation}

\begin{figure}[t]
    \centering
        \includegraphics[width=0.425\textwidth]{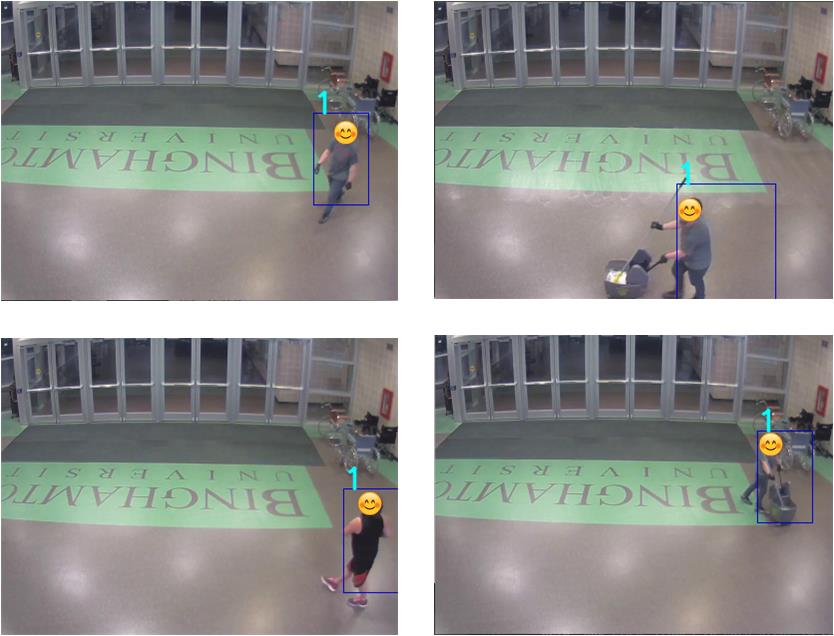}
    \caption{Visualization of the object detection and tracking.}
    \label{fig:visual}
    \vspace{-10pt}
\end{figure}

\begin{figure}[t]
    \centering
        \includegraphics[width=0.425\textwidth]{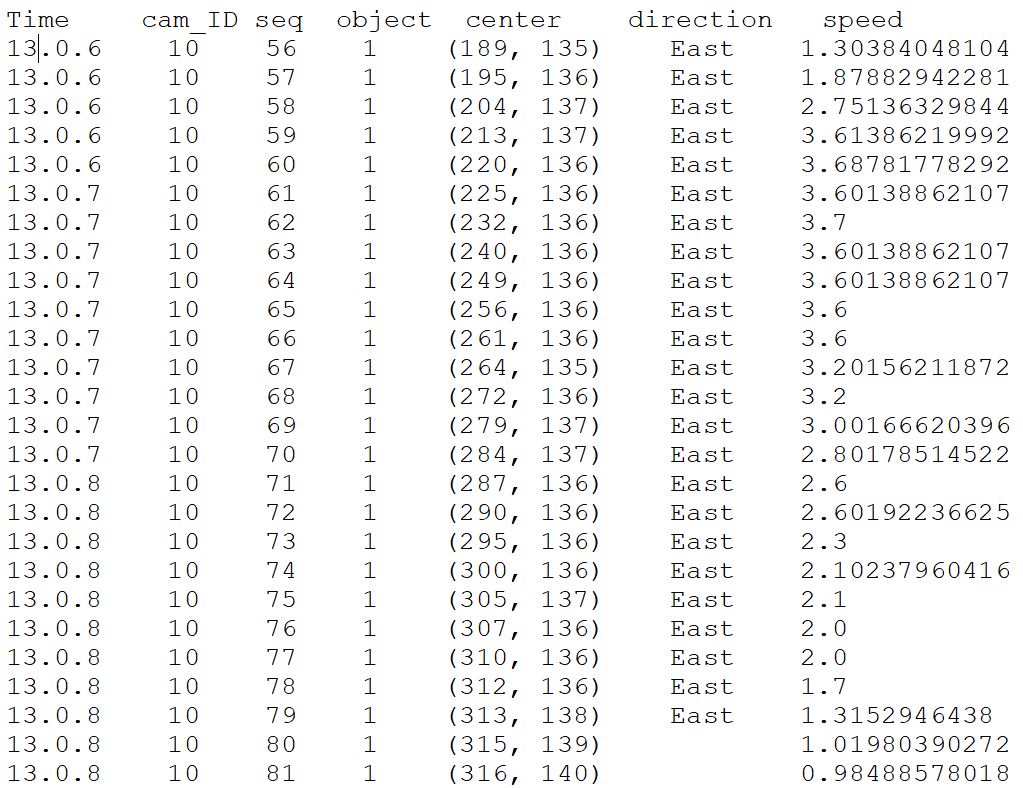}
    \caption{Contents of the feature file extracted from the live video stream.}
    \label{fig:features}
    \vspace{-10pt}
\end{figure}

Once the video is streamed to the edge device, people object detection is conducted in real-time using a light-weighted CNN \cite{nikouei2018lcnn}. The pedestrians are identified and a tracking algorithm uses the detection bounding boxes to follow the objects of interest until they exit the frame. The trackers are faster and run for each frame, while detection run only 2 times in each second. Following the object of interest will extract features based on the movement of the pedestrians in the frame. In this work, several features are considered, including their relative speed (calculated based on pixels of movement in one second and divided by the bounding box area) and direction, which is visualized in Fig. \ref{fig:visual}. The features are written in a file to be sent to the fog node. In this file, each row shows the time stamp, frame sequence number, camera ID, pedestrian ID, and then the features for that pedestrian as shown in Fig \ref{fig:features}. 

\subsubsection{Computational Overhead on Index Authentication}

\begin{figure}[t]
    \centering
        \includegraphics[width=0.425\textwidth]{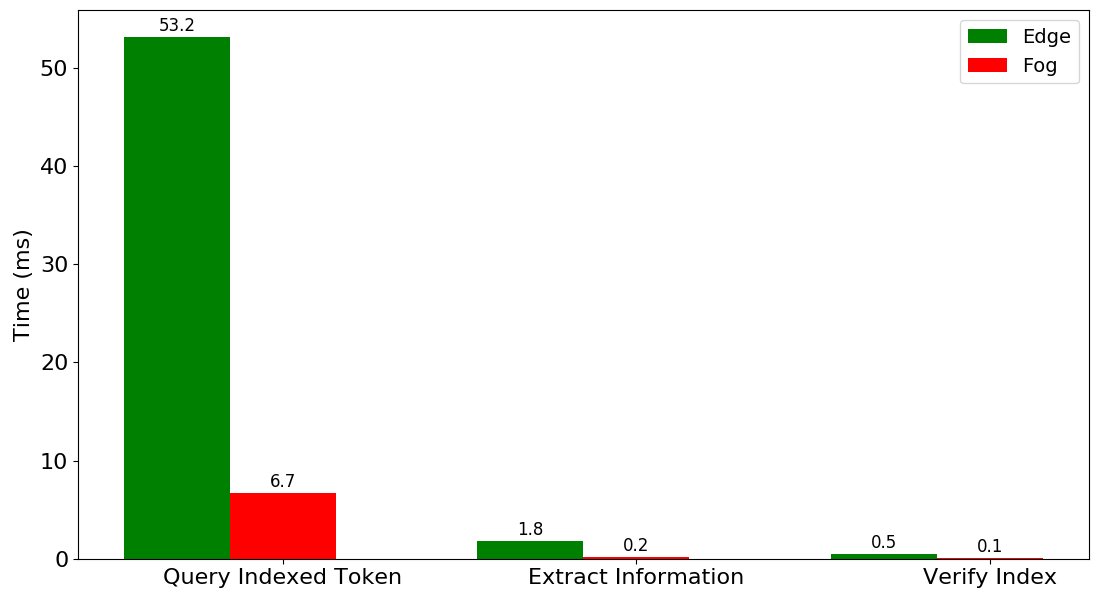}
    \caption{Computation Time for Each Stage in Index Authentication.}
    \label{fig:indexauth}
    \vspace{-10pt}
\end{figure}

The index authentication test scenarios are carried out both on the fog and edge nodes to evaluate computation overload. During the test, the average delay time is calculated against over 50 runs. As illustrated by the results shown in Fig. \ref{fig:indexauth}, query index token process, which is mainly responsible for fetching token data from the smart contract, is the most computing intensive stage among the index authorization stages. Since the fog nodes have much more computation capacity than the edge nodes do, while the execution time of querying index token on edge nodes is about 53 ms, the same operation on fog nodes only needs 6 ms. The entire index authentication process is divided into two steps, processing data in the feature files and verifying the hashed feature data. The average time of the authorization process is about 2.3 ms (1.8 ms + 0.5 ms) on edge nodes and 0.3 ms (0.2 ms + 0.1 ms) on fog nodes.

\subsubsection{Encryption Channel Analysis}

\begin{figure}[t]
    \centering
        \includegraphics[width=0.425\textwidth]{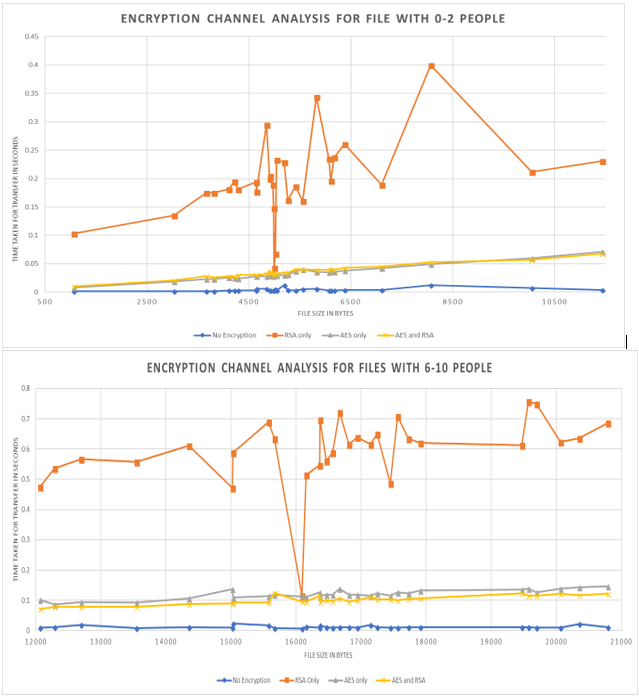}
    \caption{Encryption Channel Analysis}
    \label{fig:encrypt02}
    \vspace{-10pt}
\end{figure}

The transfer of file from fog node to the mining node through a secure channel is a necessity to avoid eavesdropping between these nodes. Attacks like man in the middle attack, ARP spoofing can cause privacy breach. Figure \ref{fig:encrypt02} represents the comparison between different encryption types to study the delay caused due to transfer from fog node to mining node. The blue line represents data transfer without any encryption, gray line represents AES encryption, orange represents RSA based encryption and yellow line represents the proposed encryption scheme with both RSA and AES encryption. It is clear that adding an additional layer of encryption does not cause more delay. The jitters caused in the data transfer speed is due to the network traffic. 

\section{Conclusions}
\label{sec:conclusion}

Technological advancements are changing the way that humans live. Nowadays many applications are moved toward automation to make it is easier for users. People object recognition focused smart surveillance helps reduce the delay in detection and rise alert timely such that the first responders are given more time to react. In order to achieve this goal with minimum delays, more computing tasks are migrated to the edge devices that are closer to the camera and abstracted features are outsourced to a fog node for anomaly detection. During this process, security concerns must be addressed to protect the data from being tampered or stolen.

In this paper we proposed a novel method to protect the indexing and feature data exchanged among nodes in a smart surveillance system that leverages the edge-fog-cloud computing paradigm. In this hierarchical architecture, 
extracted features from the video frames on the edge are sent through a secure channel to a fog node, where the features are contextualized and saved within an index table for faster and event-oriented query. The security of the communication among the nodes deployed at the edge, fog and cloud levels are provided using smart contracts that leverage the hash values of the index table to generate the next block in the blockchain network. A RSA-AES hybrid encryption method was employed to protect the extracted features of the objects of interest in the transferring from the edge nodes to the fog device before the new block is mined. After the new block is mined, the whole network will be synchronized with the new information. Using a web service, the cloud can also securely gain access to the index table and query information from one of the nodes. Experimental results have validated that the proposed method incurs very low overhead, therefore it is a promising and feasible solution for the event-oriented, real-time surveillance video query applications. 

While working on the novel authentication approach to protect the feature and indexing data, our current effort is also exploring the generation of an ideal query index database. Because it is non-trivial to identify a well-defined feature set for an efficient query indexing. One of the most significant challenges lies in the unique needs in different applications corresponding to different social and/or cultural groups. 

\section*{Acknowledgement}

Y. Chen is partially supported by the U.S. Air Force Research Laboratory (AFRL) via the Summer Faculty Fellowship Program (SFFP). The views and conclusions contained herein are those of the authors and should not be interpreted as necessarily representing the official policies, either expressed or implied, of the U.S. Government.

\ifCLASSOPTIONcaptionsoff
  \newpage
\fi



%

\bibliographystyle{IEEEtranS}

\bibliography{Index_authentic}

%





\end{document}